\documentclass[floatfix,twocolumn,showpacs,preprintnumbers,amsmath,amssymb,eqsecnum,aps]{revtex4}
\usepackage{graphicx}
\usepackage{amsthm,amsfonts}

\begin{document}

\newtheorem{teo}{Teorema}[section]

\title{Sudden death of entanglement and teleportation fidelity loss via the Unruh effect}

\author{Andr\'e G. S. Landulfo}
\email{landulfo@ift.unesp.br}
\author{George E. A. Matsas}
\email{matsas@ift.unesp.br}
\affiliation{Instituto de F\'\i sica Te\'orica, Universidade Estadual Paulista,
Rua Dr. Bento Teobaldo Ferraz, 271 - Bl. II, 01140-070, S\~ao Paulo, SP, Brazil }

\date{\today}
\begin{abstract}
We use the Unruh effect to investigate how the teleportation of quantum
states is affected when one of the entangled qubits used in the process
is under the influence of some external force.  In order to reach a
comprehensive understanding, a detailed analysis of the acceleration effect
on such entangled qubit system is performed. In particular, we calculate the
mutual information and concurrence between the two qubits and show that the
latter has a ``sudden death" at a finite acceleration, whose value will
depend on the time interval along which the detector is accelerated.
\end{abstract}

\maketitle

\section{Introduction}
\label{sec:Introduction}

The teleportation of quantum states is undoubtedly one of the most
interesting effects unveiled in the last decade. In the original work by
Bennett et al~\cite{Betal93} the system is considered to be isolated
from external forces and the maximally entangled qubit pair is unitarily
evolved. As a natural development, Alsing and Milburn analyzed the
case when the system is not quite isolated~\cite{AM03}. In their set
up, Bob is replaced by a uniformly accelerated observer named Rob.
Alice and Rob each hold an optical cavity at rest in their local frames,
which are assumed to be initially free of Minkowski photons. Each cavity
supports two orthogonal Minkowski modes $A_i$ and $R_i$ $(i=1,2)$ with
the same frequency, where hereafter $A$ and $R$ will be used to label
Alice and Rob, respectively. At the moment that Alice and Rob overlap,
they create an entangled pair
\begin{equation}
|{\bf 0} \rangle_M \otimes |{\bf 0} \rangle_M +
|{\bf 1} \rangle_M \otimes |{\bf 1} \rangle_M
\label{qubitAM}
\end{equation}
where
$$
|{\bf 0} \rangle_M = |1 \rangle_{X_1} \otimes | 0 \rangle_{X_2},
\;\;\;\;
|{\bf 1} \rangle_M = |0 \rangle_{X_1} \otimes | 1 \rangle_{X_2}
$$
and $X=A$ and $R$ for the first and second qubits in
Eq.~(\ref{qubitAM}), respectively. Then it is argued that
as Rob is accelerated, his cavity would be populated by
thermally excited Rindler photons as it would be predicted
by the Unruh effect~\cite{U76} (see also Ref.~\cite{CHM08} for
a recent review on the Unruh effect and its applications) and
is concluded that the teleportation fidelity would be reduced.
We note however, that the set up proposed above presents
some conceptual difficulties~\cite{SU05}. In particular, the
relationship between the Minkowski and Rindler modes as
used in Ref.~\cite{AM03} is valid in the Minkowski spacetime
without boundary conditions imposed by the presence of
cavities.

In order to circumvent these difficulties, we introduce here
a distinct set up which avoids the use of cavities. For this
purpose the qubits are modeled by a two-level semiclassical
detector coupled to a massless scalar field. The detector is
classical in the sense that it has a well defined worldline
but quantum because of the nature of its internal degrees of
freedom. The paper is organized as follows. In Sec.~\ref{sec:detector}
we introduce our qubit and its interaction with the Klein-Gordon
field. In Sec.~\ref{sec:entanglement} we entangle a pair of those
qubits and use the Unruh effect to calculate its final state when
one of them is uniformly accelerated for some fixed amount of
proper time while the other one is inertial. Next, we investigate
the corresponding mutual information and concurrence as a function
of the non-inertial qubit acceleration. In particular, we verify
that the qubit system experiences a sudden death of entanglement
(see, e.g., Ref.~\cite{Aetal07,DH09} and references therein) at a
finite proper acceleration. In Sec.~\ref{sec:teleportation} we
revisit the original teleportation protocol~\cite{Betal93} when the
inertial Bob is replaced by the accelerated Rob and calculate how
the the teleportation fidelity diminishes as the acceleration grows.
We dedicate Sec.~\ref{sec:conclusions} for our closing remarks and
to establish a relationship between our theoretical model and a
possible experimental physical set up. We adopt spacetime signature
$(- + + +)$ and assume natural units $c=\hbar=1$ unless stated
otherwise.

\section{Two-level detector qubit model}
\label{sec:detector}

We model our qubit in Minkowski spacetime $( \mathbb{R}^4 , g_{ab} )$
through a two-level detector
with energy gap $\Omega$ as introduced by Unruh and Wald~\cite{UW84}.
Here, we briefly revisit the corresponding detector theory and derive the
necessary results for the forthcoming sections.
The detector proper Hamiltonian is defined as
\begin{equation}
H_D=\Omega D^{\dagger}D,
\label{detetor}
\end{equation}
where
$D |0\rangle = D^{\dagger} |1\rangle = 0$,
$D |1\rangle=|0\rangle$
and
$D^{\dagger}|0\rangle=|1\rangle$,
and
$ |0\rangle, |1\rangle$ are the corresponding unexcited, excited energy
eigenstates, respectively. We couple the detector to a massless scalar
field $\phi$ satisfying the Klein-Gordon equation~\cite{indeces}
\begin{equation}
\nabla_a \nabla^a \phi=0
\label{KG}
\end{equation}
through the Hamiltonian
\begin{equation}
H_{\rm int}(t)=
\epsilon(t) \int_{\Sigma_t} d^3 {\bf x} \sqrt{-g} \phi(x) [\psi({\bf x})D +
                           \overline{\psi}({\bf x})D^{\dagger}],
\label{int}
\end{equation}
where
$\phi (x)$ is the free Klein-Gordon field operator,
$g\equiv {\rm det} (g_{ab}) $ and ${\bf x}$ are coordinates defined on the
Cauchy surface $\Sigma_{t = {\rm const}}$ associated with some suitable timelike
isometry. For our present purposes we assume that the detector follows either the
inertial or the uniformly accelerated isometry of the Minkowski spacetime.
Here $\epsilon \in C^{\infty}_0(\mathbb{R})$ is a smooth
compact-support real-valued function, which keeps the detector switched on for a
finite amount of proper time $\Delta$ (for more details on finite-time detectors
see, e.g., Ref.~\cite{HMP93}) and $\psi \in C^{\mathbb{C} \infty}_0(\Sigma_t)$
is a smooth compact-support complex-valued function, which models the  fact that the
detector only interacts with the field in a neighborhood of its worldline.
The same detector model was recently used by Kok and Yurtsever to analyze the
decoherence of an accelerated qubit due to the Unruh effect~\cite{KY03}.
Using Eqs.~(\ref{detetor})-(\ref{int}) we cast the total Hamiltonian as
\begin{equation}
H_{D\, \phi}= H_0 + H_{\rm int},
\end{equation}
where $H_0 = H_D + H_{KG}$ is the combined detector-field free Hamiltonian.
In the interaction picture the state $|\Psi^{D \phi}_t \rangle$ describing
the system at moment $t$ can be written as
\begin{equation}
|\Psi^{D \phi}_t \rangle =
T \exp[-i\int_{-\infty}^t dt' H_{\rm int} ^I(t')] |\Psi^{D \phi}_{-\infty} \rangle,
\label{Dyson1}
\end{equation}
where $T$ is the time-ordering operator and
\begin{equation}
H_{\rm int}^I (t) = U^{\dagger}_0(t) H_{\rm int} (t) U_0 (t)
\end{equation}
with $U_0 (t)$ being the unitary evolution operator associated with
$H_0 (t)$. By using Eq.~(\ref{Dyson1}), we write
$|\Psi^{D \phi}_{\infty} \rangle = |\Psi^{D \phi}_{t > \Delta} \rangle$
as
\begin{equation}
|\Psi^{D \phi}_{\infty} \rangle =
T \exp [-i\int d^4x\sqrt{-g}\phi(x) (fD + \overline{f}D^{\dagger})] |\Psi^{D \phi}_{-\infty} \rangle,
\label{Dyson2}
\end{equation}
where
$f \equiv \epsilon(t) e^{-i\Omega t}\psi ({\bf x})$
is a compact support complex function defined in Minkowski spacetime
and we have used that
$D^I=e^{-i\Omega t}D$.
In first perturbation order, Eq.~(\ref{Dyson2}) becomes
\begin{equation}
|\Psi^{D \phi}_{\infty} \rangle
= [I - i(\phi(f)D + \phi(f)^{\dagger} D^{\dagger}) ] |\Psi^{D \phi}_{-\infty} \rangle,
\label{primeira_ordem}
\end{equation}
where~\cite{wald94}
\begin{eqnarray}
\phi(f) &\equiv& \int d^4x \sqrt{-g } \phi(x) f
\nonumber\\
        & = & i [a(\overline{KE\overline{f}})-a^{\dagger}(KEf)]
\label{phi(f)}
\end{eqnarray}
is an operator valued distribution obtained by smearing out the field operator
by the testing function $f$ above.
Here $a(\overline{u})$ and $a^{\dagger} (u)$ are annihilation and creation
operators of $u$ modes, respectively, the $K$ operator takes the positive-frequency
part of the solutions of Eq.~(\ref{KG}) with respect to the timelike isometry,
and
\begin{equation}
Ef = \int d^4x'\sqrt{-g(x')} [G^{\rm adv}(x, x')-G^{\rm ret}(x, x')] f(x'),
\label{Ef}
\end{equation}
where
$G^{\rm adv}$ and $ G^{\rm ret}$ are the advanced and retarded Green functions, respectively.
Next, by imposing that $\epsilon(t)$ is a very slow-varying function of time
compared to the frequency $\Omega$ and that $\Delta \gg \Omega^{-1}$,
we have that $f$ is an approximately positive-frequency function, i.e.,
$KEf\approx Ef$ and $KE\overline{f}\approx 0$ (see appendix~\ref{appendix}). Now, by defining
\begin{equation}
\lambda \equiv -KEf,
\label{aux1}
\end{equation}
we cast Eq.~(\ref{phi(f)}) as
\begin{equation}
\phi(f)\approx i a^{\dagger}(\lambda)
\label{phi(f)2}
\end{equation}
and Eq.~(\ref{primeira_ordem}) as
\begin{equation}
|\Psi^{D \phi}_{\infty} \rangle=
(I + a^{\dagger}(\lambda)D -  a(\overline{\lambda})D^{\dagger} ) |\Psi^{D \phi}_{-\infty} \rangle.
\label{primeira_ordem_2}
\end{equation}
The expression above carries the well known physical message that the excitation and
deexcitation of an Unruh-DeWitt detector following a timelike isometry is associated
with the absorption and emission, respectively, of a particle as ``naturally"  defined
by observers comoving with the detector, i.e., in our case, Minkowski and Rindler
particles for inertial and uniformly accelerated observers, respectively.

\section{Entangled qubit pair and the Unruh effect}
\label{sec:entanglement}

Let us consider now a two-qubit system initially entangled as given by
\begin{equation}
|\Psi_{AR}\rangle = \alpha |0_A\rangle \otimes |1_R\rangle + \beta|1_A\rangle \otimes |0_R\rangle
\label{initialqubit}
\end{equation}
with
$|\alpha |^2 + | \beta |^2 = 1$,
where
$\{|0_X\rangle, |1_X\rangle \}$
is an orthonormal basis of the internal qubit space
$\mathfrak{H}_X$ and $X=A, R$. The free Hamiltonian for each one of the
detectors is given by Eq.~(\ref{detetor}) with $D$ replaced by $A$ or $R$
depending on the detector. Now, we impose that Alice's detector
is kept inertial in contrast to Rob's one which is uniformly
accelerated for a finite proper time $\Delta$, having worldline
\begin{equation}
   t(\tau ) = a^{-1}\sinh{a\tau},\,  x(\tau ) = a^{-1}\cosh{a\tau}, \,  y(\tau) =z (\tau) =0,
\end{equation}
where $\tau$ and $a$ are the detector's proper time and acceleration,
respectively, and here $(t,x,y,z)$ are the usual Cartesian
coordinates of Minkowski spacetime. The detectors are designed to be switched on only
when they are accelerated. Thus, Alice's inertial qubit only interacts  with the
scalar field indirectly through Rob's detector. At the end of the paper we
discuss a laboratory situation which realizes these assumptions.

Rob's qubit interacts with the field according to the Hamiltonian~(\ref{int})
with the proper replacements: $D \to R$ and $t \to \tau$, where $\Sigma_\tau$
are spacelike hypersurfaces orthogonal to the congruence of boost isometries
to which Rob's detector worldline belongs. The total Hamiltonian
is given by
\begin{equation}
H_{A\, R\, \phi} = H_A + H_R + H_{KG} +  H_{\rm int}.
\end{equation}
The corresponding Hilbert space associated with our system can be written now as
$\mathfrak{H}_T = \mathfrak{H}_A \otimes \mathfrak{H}_R \otimes
\mathfrak{F}_s (\mathfrak{H}_I\oplus \mathfrak{H}_{II})$,
where $\mathfrak{F}_s (\mathfrak{H}_I\oplus \mathfrak{H}_{II}) $ is the
symmetric Fock space of $\mathfrak{H}_I\oplus \mathfrak{H}_{II}$ with
$\mathfrak{H}_I$ being the Hilbert space of positive-frequency
solutions with respect to $\tau$ with initial data on $\Sigma_I$ which
is the portion of $\Sigma_{\tau=0}$ in the right Rindler wedge  defined by
$x> |t|$, and analogously for  $\mathfrak{H}_{II}$ and the left Rindler
wedge defined by $x< -|t|$.

Next by using the fact that Rob's detector is the only one which
interacts with the field and that this is confined in the right Rindler
wedge, we use Eq.~(\ref{primeira_ordem_2}) to evolve our initial state
\begin{equation}
|\Psi^{AR \phi}_{-\infty} \rangle = |\Psi_{AR}\rangle \otimes |0_M\rangle,
\label{initialstate}
\end{equation}
with  $|0_M\rangle$ being the Minkowski vacuum (i.e., the no-particle
state as defined by inertial observers), to its asymptotic form
\begin{equation}
|\Psi^{AR \phi}_{\infty} \rangle=
(I + a_{R I}^{\dagger}(\lambda)R
   - a_{R I}(\overline{\lambda}) R^{\dagger} ) |\Psi^{AR \phi}_{-\infty} \rangle,
\label{evolution}
\end{equation}
where the labels in $a_{R I}^{\dagger}$ and $a_{R I}$ emphasize that
they are creation  and annihilation operators of  Rindler
modes in the right wedge ($I$), $\lambda = -KEf \approx Ef $, and here
$f = \epsilon(\tau) e^{-i\Omega \tau} \psi({\bf x})$.
By using Eqs.~(\ref{initialqubit}) and~(\ref{initialstate})
in Eq.~(\ref{evolution}), we obtain
\begin{eqnarray}
| \Psi^{AR \phi}_{\infty}\rangle
& = &
|\Psi^{AR \phi}_{-\infty} \rangle
 + \alpha |0_A\rangle \otimes |0_R\rangle
 \otimes(a_{R I}^{\dagger}(\lambda)|0_M\rangle)
 \nonumber \\
& + & \beta|1_A\rangle \otimes |1_R\rangle\otimes(a_{R I}(\overline{\lambda})|0_M\rangle) .
\label{evolutionAUX}
\end{eqnarray}
In order to proceed, we write $a_{R I}$ and $a^{\dagger}_{R I}$
in terms of the annihilation, $a_M$, and creation, $a_M ^{\dagger}$,
operators of Minkowski modes as~\cite{UW84}
\begin{eqnarray}
a_{R I}(\overline{\lambda})&=&
\frac{a_M(\overline{F_{1 \Omega}})+
e^{-\pi \Omega/a} a_M ^{\dagger} (F_{2 \Omega})}{(1- e^{-2\pi\Omega/a})^{{1}/{2}}},
\label{aniq} \\
a^{\dagger}_{R I}(\lambda)&=&
\frac{a^{\dagger}_M (F_{1 \Omega}) +
e^{-\pi \Omega/a}a_M(\overline{F_{2 \Omega}})}{(1- e^{-2\pi\Omega/a})^{{1}/{2}}}
\label{cria},
\end{eqnarray}
where
\begin{eqnarray}
F_{1 \Omega}&=&
\frac{\lambda+ e^{-\pi\Omega/a}\lambda\circ w}{(1- e^{-2\pi\Omega/a})^{{1}/{2}}},
\\
F_{2 \Omega}&=&
\frac{\overline{\lambda\circ w}+ e^{-\pi\Omega/a}\overline{\lambda}}{(1- e^{-2\pi\Omega/a})^{{1}/{2}}},
\end{eqnarray}
$w(t, x, y, z)=(-t, -x, y, z)$
is the wedge reflection isometry,
and we recall that whenever
$\varphi\in \mathfrak{H}_I$
then
$\varphi\circ w \in \overline{\mathfrak{H}}_{II}$.
For further convenience, let us define from Eq.~(\ref{aux1})
\begin{equation}
\nu^2 \equiv ||\lambda||^2,
\label{nulambda}
\end{equation}
where
\begin{equation}
(F_{i \Omega}, F_{j \Omega})_{KG} = ||\lambda||^2 \delta_{i j} , \;\; i \in \{1, 2 \}.
\label{KGF}
\end{equation}
Here we write the Klein-Gordon internal product
$$
( F_{i \Omega}, F_{j \Omega})_{KG}
\equiv i \int_\Sigma d^3 x \sqrt{h} (\overline F_{i \Omega} \nabla_a F_{j \Omega}
- (\nabla_a {\overline F_{i \Omega}}) F_{j \Omega}) n^a
$$
between the positive-\-frequency solutions,
$F_{i \Omega}$ and $F_{j \Omega}$, with respect to the Minkowski time $t$
taken on a Cauchy surface $\Sigma$ with unit orthogonal vector $n^a$ and
$h \equiv {\rm det} (h_{ab})$ with $h_{ab}$ being the restriction of $g_{ab}$ on $\Sigma$.
By assuming our detector to be localized as given by the Gaussian
$\psi(\mathbf{x})= (\kappa\sqrt{2\pi})^{-3} \exp (-\mathbf{x}^2/2\kappa^2)$
with variance $\kappa = {\rm const} \ll 1$, we show in the appendix~\ref{appendix} that
\begin{equation}
\nu^2=\frac{\epsilon^2 \Omega\Delta}{2\pi}e^{-\Omega^2 \kappa^2}.
\label{nu}
\end{equation}
\begin{figure}[t]
\begin{center}
\includegraphics[height=0.22\textheight]{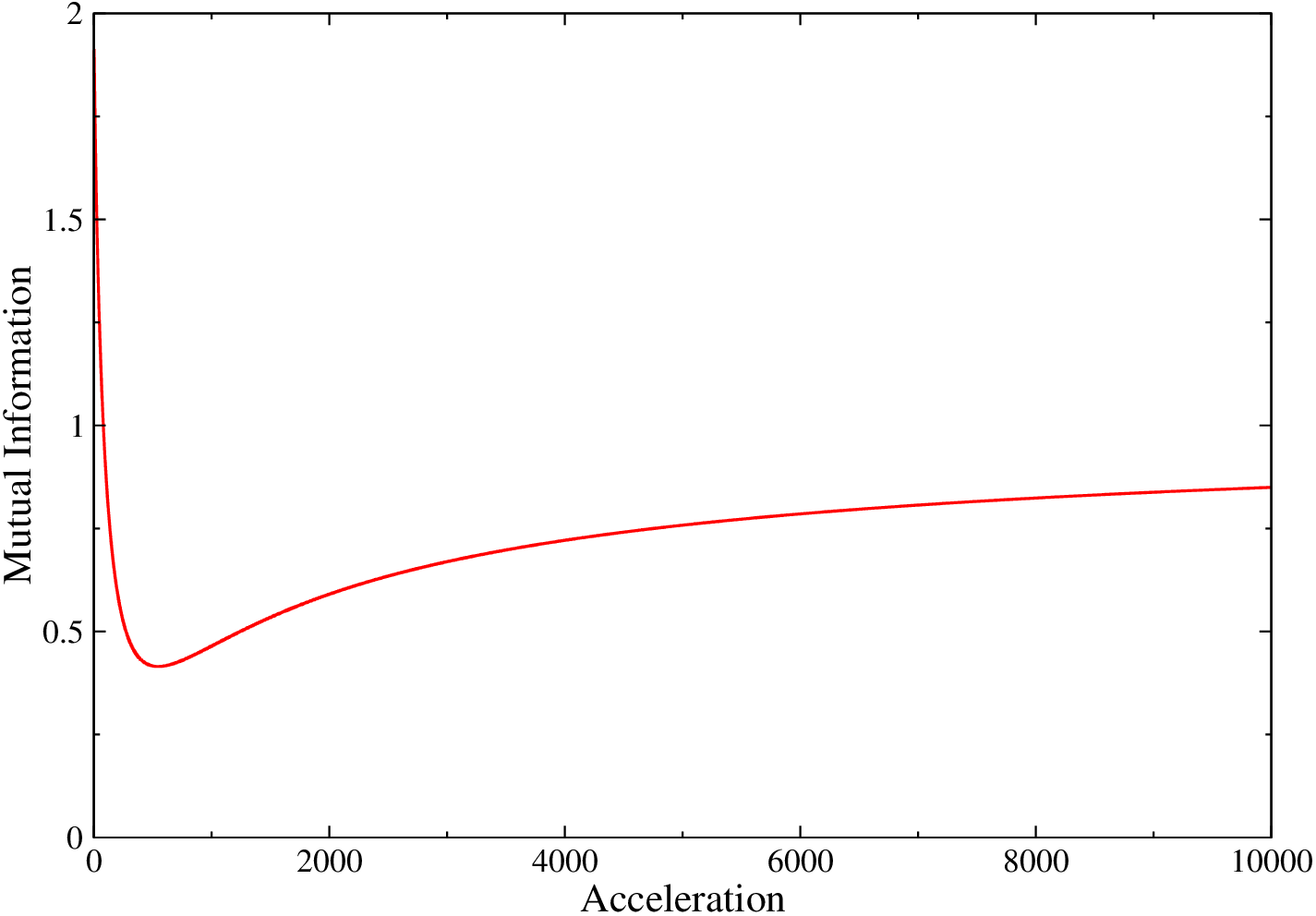}
\caption{The graph exhibits the mutual information $I(A:R)$
for a singlet initial state as a function of
acceleration $ a/\Omega$, with $\epsilon^2 = 8\pi^2 \cdot 10^{-6}$,
$\Omega=100$, $\Delta=1000$ and $\kappa=0.02$.
The most interesting feature is related with the fact that
the curve is not monotonic, acquiring its minimum value
at $a_0/\Omega\approx  545.75$. We note that
the dimensionless quantity $a/\Omega$ reflects  the temperature of
the Unruh thermal bath as experienced by Rob's detector
per its energy gap (up to a $1/(2 \pi)$ factor).}
\label{Fig1_MI}
\end{center}
\end{figure}

Now, by using Eqs.~(\ref{aniq}) and~(\ref{cria}) to write
\begin{eqnarray}
a_{R I}(\overline{\lambda})|0_M\rangle
&=&\frac{\nu e^{-\pi \Omega/a}}{(1-e^{-2\pi \Omega/a})^{{1}/{2}}}|1_{\tilde{F}_{2\Omega}}\rangle,\\
a_{R I}^{\dagger}(\lambda)|0_M\rangle
&=&\frac{\nu}{(1-e^{-2\pi \Omega/a})^{{1}/{2}}}|1_{\tilde{F}_{1\Omega}}\rangle ,
\end{eqnarray}
we cast Eq.~(\ref{evolutionAUX}) in the form
\begin{eqnarray}
| \Psi^{AR \phi}_{\infty} \rangle &= &
| \Psi^{AR \phi}_{-\infty} \rangle  +
                            \alpha \nu \frac{|0_A\rangle
\otimes |0_R \rangle \otimes|1_{\tilde{F}_{1 \Omega}}\rangle}{(1- e^{-2\pi\Omega/a})^{{1}/{2}}}
\nonumber \\
                         &+& \beta \nu e^{-\pi\Omega/a}\frac{|1_A\rangle
\otimes |1_R\rangle\otimes|1_{\tilde{F}_{2 \Omega}}\rangle}{(1- e^{-2\pi\Omega/a})^{{1}/{2}}},
\label{Dyson3}
\end{eqnarray}
where
$\tilde{F}_{i \Omega}=F_{i \Omega}/\nu$. Notice that the fact that every Rob's qubit
transition demands the emission of a Minkowski particle is codified in Eq.~(\ref{Dyson3}).

The density matrix which describes the two-qubit state is obtained tracing out the
scalar field degrees of freedom, namely,
\begin{equation}
\rho_{\infty}^{AR} =
||\Psi^{AR \phi}_{\infty}||^{-2}
{\rm tr}_{\phi}|\Psi^{AR \phi}_{\infty}\rangle \langle \Psi^{AR \phi}_{\infty}|,
\label{rhof1}
\end{equation}
where
$$
||\Psi^{AR \phi}_{\infty}||^2 = 1+\frac{|\alpha|^2 \nu^2}{1-e^{-2\pi\Omega/a}}
+ \frac{|\beta|^2 \nu^2 e^{-2\pi\Omega/a}}{1-e^{-2\pi\Omega/a}}
$$
normalizes the final density matrix, i.e., ${\rm tr} \rho_{\infty}^{AR} = 1$. By working out
Eq.~(\ref{rhof1}), we obtain
\begin{eqnarray}
\rho_{\infty}^{AR} &=& 2 S^{\alpha \beta}_0|\Psi_{AR}\rangle \langle \Psi_{AR}|
           + S^{\alpha \beta}_2|0_A\rangle\otimes|0_R\rangle \langle 0_A|\otimes \langle 0_R|
           \nonumber \\
           &+& S^{\alpha \beta}_1|1_A\rangle\otimes|1_R\rangle \langle 1_A|\otimes \langle 1_R|,
\label{rhof2}
\end{eqnarray}
where
\begin{eqnarray}
S^{\alpha \beta}_0 &=&
\frac{(1-e^{-2\pi\Omega/a})/2}{(1-e^{-2\pi\Omega/a})
+|\alpha|^2 \nu^2 + |\beta|^2 \nu^2 e^{-2\pi\Omega/a}},
\label{s0}
\nonumber \\
S^{\alpha \beta}_1 &=&
\frac{|\beta|^2 \nu^2 e^{-2\pi\Omega/a}}{(1-e^{-2\pi\Omega/a})
+ |\alpha|^2 \nu^2 +|\beta|^2 \nu^2 e^{-2\pi\Omega/a}},
\label{s1}
\nonumber \\
S^{\alpha \beta}_2&=& \frac{|\alpha|^2 \nu^2} {(1-e^{-2\pi\Omega/a})
+ |\alpha|^2\nu^2 +|\beta|^2 \nu^2 e^{-2\pi\Omega/a}},
\label{s2}
\nonumber
\end{eqnarray}
and we verify that $2S^{\alpha \beta}_0+S^{\alpha \beta}_1+S^{\alpha \beta}_2=1$.
For the sake of convenience, we cast Eq.~(\ref{rhof2}) in matrix form as
\begin{equation}
\rho_{\infty}^{AR}=\left( \begin{array}{c c c c}
S^{\alpha \beta}_2 \, & 0 \, & 0 \, & 0 \, \\
 0 \, & 2 \, |\alpha|^2\, S^{\alpha \beta }_0 \, & 2\,\alpha\, \overline{\beta}\, S^{\alpha \beta}_0 \, & 0\, \\
 0 \, & 2 \, \overline{\alpha}\, \beta\, S^{\alpha \beta}_0 \, & 2\, |\beta|^2\, S^{\alpha \beta}_0  \, & 0\, \\
 0 \, & 0 \,                                     & 0 \, & S^{\alpha \beta}_1 \,
\end{array}
\right),
 \label{estadofinal}
\end{equation}
where we have used the basis
$$
\{|0_A \rangle\otimes |0_R\rangle,
|0_A\rangle\otimes |1_R\rangle,
|1_A\rangle \otimes|0_R\rangle,
|1_A\rangle \otimes|1_R\rangle\}.
$$
\begin{figure}
\begin{center}
\includegraphics[height=0.22\textheight]{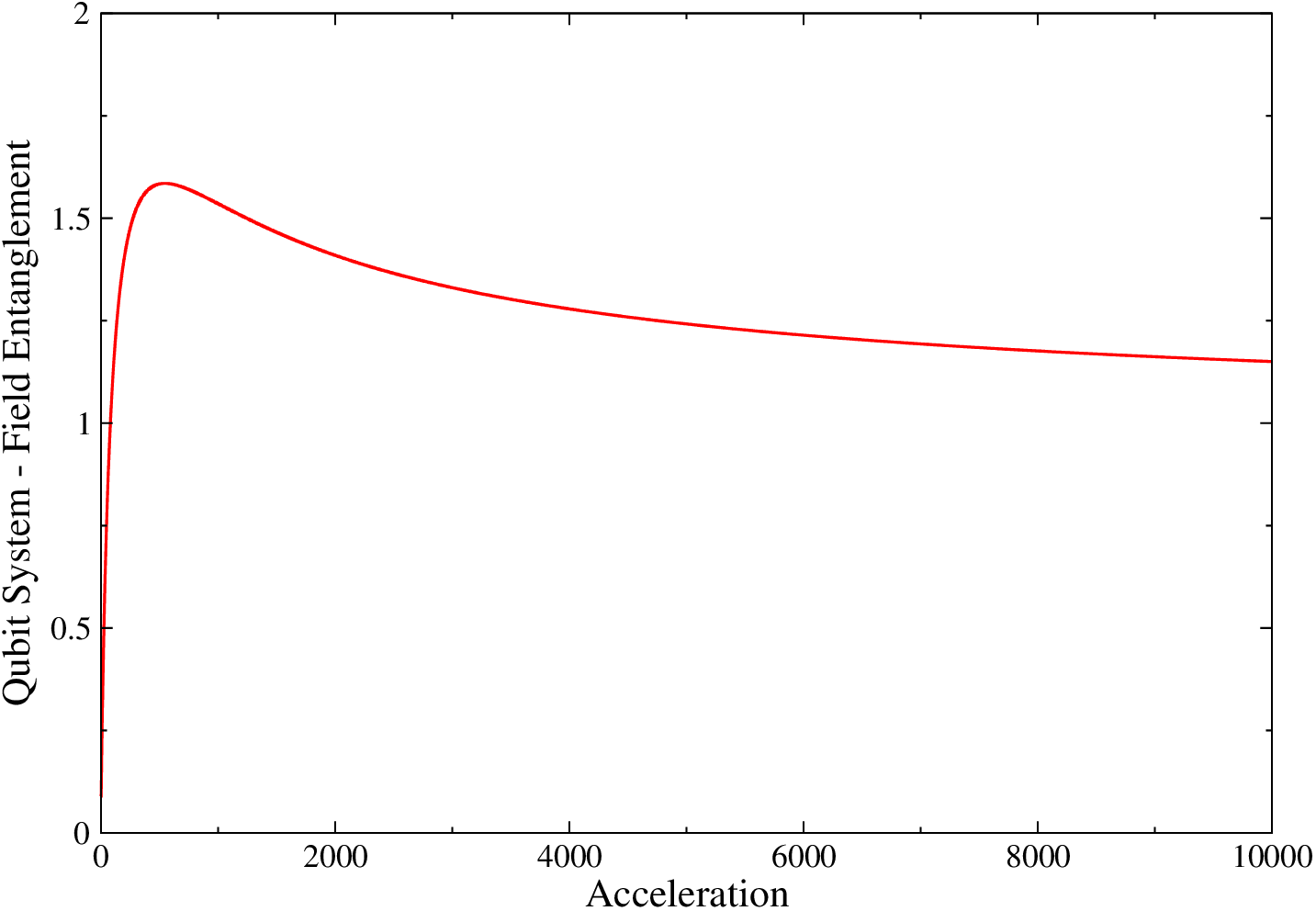}
\caption{The graph shows the entanglement, $E^{AR \phi}$, between the
qubit system and field as a function of acceleration $a/\Omega$
assuming the same initial state and $\epsilon$, $\Omega$, $\Delta$, and $\kappa$
parameters as in Fig.~\ref{Fig1_MI}. We note that the entanglement takes its
maximum value, $E^{AR \phi}_{\rm max} \approx 1.58$, at
$a_0/\Omega\approx 545.75$, precisely where $I(A:R)$ has its minimum.
This is interesting to note that because the normalized
$| \Psi^{AR \phi}_\infty \rangle$ is  Schimidt decomposed~\cite{A07}
[see Eq.~(\ref{Dyson3})], the corresponding
Schimidt number is $3$ and the maximum entanglement
$ E^{AR \phi}_{\rm max} =\log 3 \approx 1.58$.}
\label{Fig2_ent}
\end{center}
\end{figure}

\subsection{Mutual information}
\label{subsec:mutualinformation}

In order to extract information on the correlation
between the qubits $A$ and $R$, we calculate the mutual
information~\cite{A07,NC00}
\begin{equation}
I(A:R) = S(\rho_{\infty}^A) + S(\rho_{\infty}^R) - S(\rho_{\infty}^{AR}),
\end{equation}
where
$ 0 \leq I(A:R) \leq 2$.
Here
$\rho_{\infty}^A ={\rm tr}_R \, \rho_{\infty}^{AR}$,
$\rho_{\infty}^R= {\rm tr}_A \, \rho_{\infty}^{AR}$,
and $S (\rho) = - {\rm tr}\, (\rho \log_2 \rho)$ is the von Neumann
entropy. In Fig.~\ref{Fig1_MI} we plot the mutual information
for a fixed proper time interval $\Delta$ along which Rob's detector is
accelerated, assuming the two-qubit system to be initially in a singlet
state: $\alpha=-\beta=1/\sqrt{2}$.
We see that for low enough accelerations, the mutual information
keeps its value close to the maximum one, $I(A:R) \approx 2$,
as expected. This is so because for very low accelerations the
temperature of the Unruh thermal bath is small containing, thus,
quite few particles with proper energy $\Omega$ able to interact
with the detector.
The reason why $ I(A:R) \neq 2$ for arbitrarily small $a$ is because
even inertial detectors have a non-zero probability of spontaneously
decaying with the emission of a Minkowski particle, which carries away
information from the qubit-system. For arbitrarily large accelerations,
where the detector experiences high Unruh temperatures, we have
$I(A:R) \to 1$, indicating that the qubits are still correlated but not
entangled, as it can be seen directly from Eq.~(\ref{rhof2}):
$$
\rho_{\infty}^{AR} \stackrel{a\to \infty}{\longrightarrow}
\frac{1}{2}|0_A\rangle\otimes|0_R\rangle \langle 0_A|\otimes \langle 0_R|
+ \frac{1}{2}|1_A\rangle\otimes|1_R\rangle \langle 1_A|\otimes \langle 1_R|.
$$

\begin{figure}
\begin{center}
\includegraphics[height=0.22\textheight]{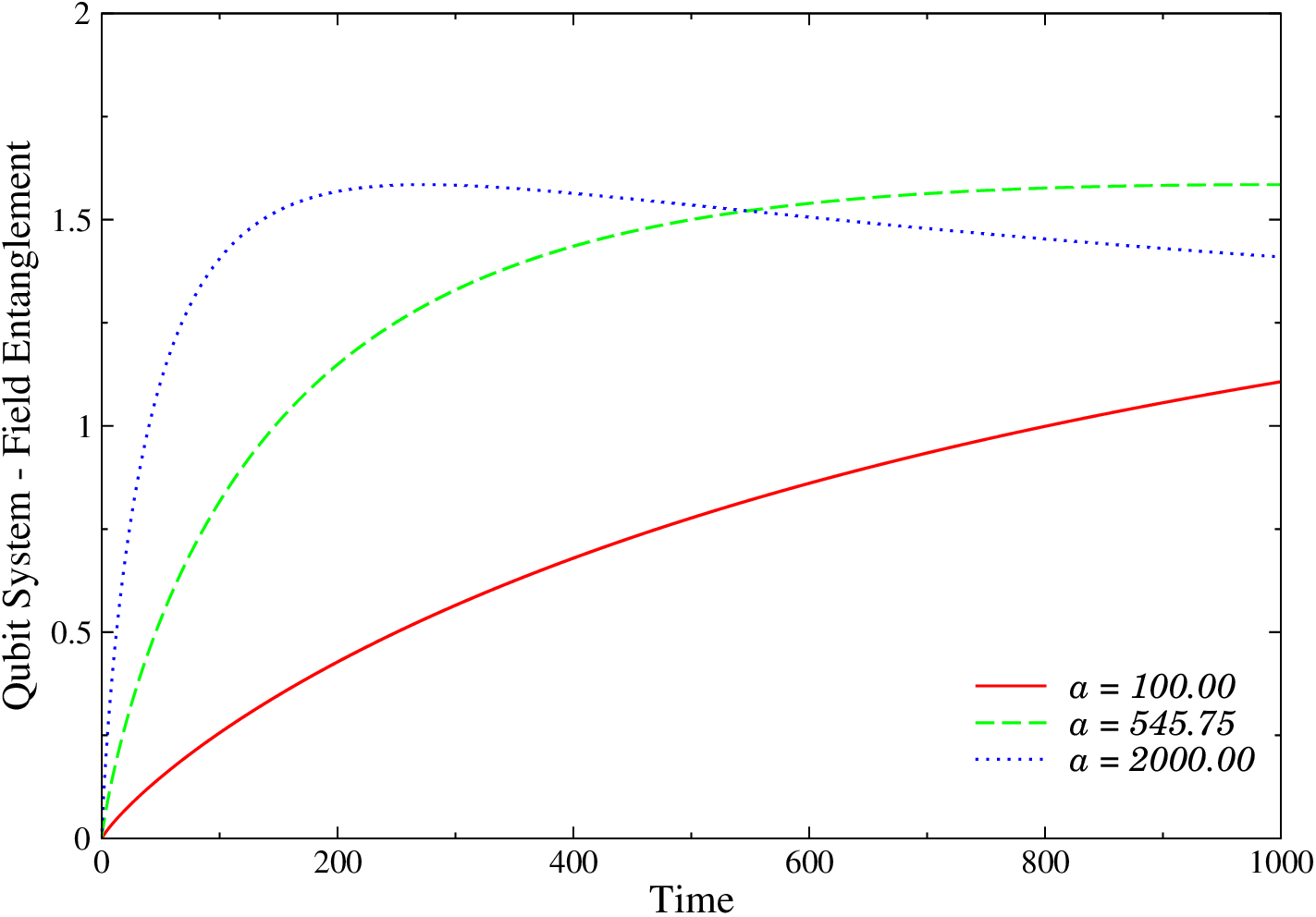}
\caption{The graph follows the behavior of the entanglement,
$E^{AR \phi}$, along the time for three distinct proper
accelerations: $a/\Omega= 100$ (full line), $a/\Omega= a_0= 545.75$ (dashed line)
and $a/\Omega = 2000$ (dotted line) assuming the same initial state
and $\epsilon$, $\Omega$, $\Delta$, and $\kappa$ parameters as in Fig.~\ref{Fig1_MI}.
We note that for $a\leq a_0$,
$E^{AR \phi}$ increases monotonically as a function of time. However,
for $a> a_0$, the qubit system and field get maximally entangled
at some time $\tau= \tau_e < \Delta$, after which the qubit system
recovers back part of its correlations from the total system.
Although not being visually evident, the graph is plotted in the acceleration
time interval $\tau = [1, \Delta]$, which respects the constraint
$\Omega \tau \gg 1$ [see discussion above Eq.~(\ref{aux1})].}
\label{Fig3_time}
\end{center}
\end{figure}

In order to get a better understanding of the physical content codified
in Fig.~\ref{Fig1_MI}, this is interesting to analyze the entanglement
between the two-qubit system and the field. Since $|\Psi^{AR \phi}_\infty \rangle$
is a pure state, the entanglement between the qubits and the field is given
by~\cite{A07}
\begin{equation}
E^{AR \phi} = S(\rho_{\infty}^{AR})=S(\rho_{\infty}^{\phi}),
\end{equation}
where $\rho_{\infty}^{AR}$ was defined in Eq.~(\ref{rhof1}) and  $\rho_{\infty}^{\phi}$ is the density matrix
obtained analogously by taking the partial trace on the qubits degrees of freedom.
In Fig.~\ref{Fig2_ent} we plot the qubit system-field entanglement for the situation described in
Fig.~\ref{Fig1_MI}. The qubit system-field entanglement $E^{AR \phi}$ is small  for low enough
accelerations, since $|\Psi^{AR \phi}_\infty \rangle$ is approximately separable (but not exactly
separable because again of the non-zero probability of spontaneous deexcitation of inertial
detectors) in contrast to the case of arbitrarily large accelerations where  $E^{AR \phi}$
approaches the unity. As for the mutual information, the qubit system-field entanglement
has a non-trivial behavior acquiring its maximum value at $a= a_0$, which is
precisely where $I(A:R)$ has its minimum (see Fig.~\ref{Fig1_MI}). For $a \gtrless a_0$, the qubit
system recovers part of its correlations after some time $\tau = \tau_e \lessgtr \Delta$
as shown in Fig.~\ref{Fig3_time}.

\subsection{Concurrence}
\label{subsec:concurrence}

\begin{figure}[t]
\begin{center}
\includegraphics[height=0.22\textheight]{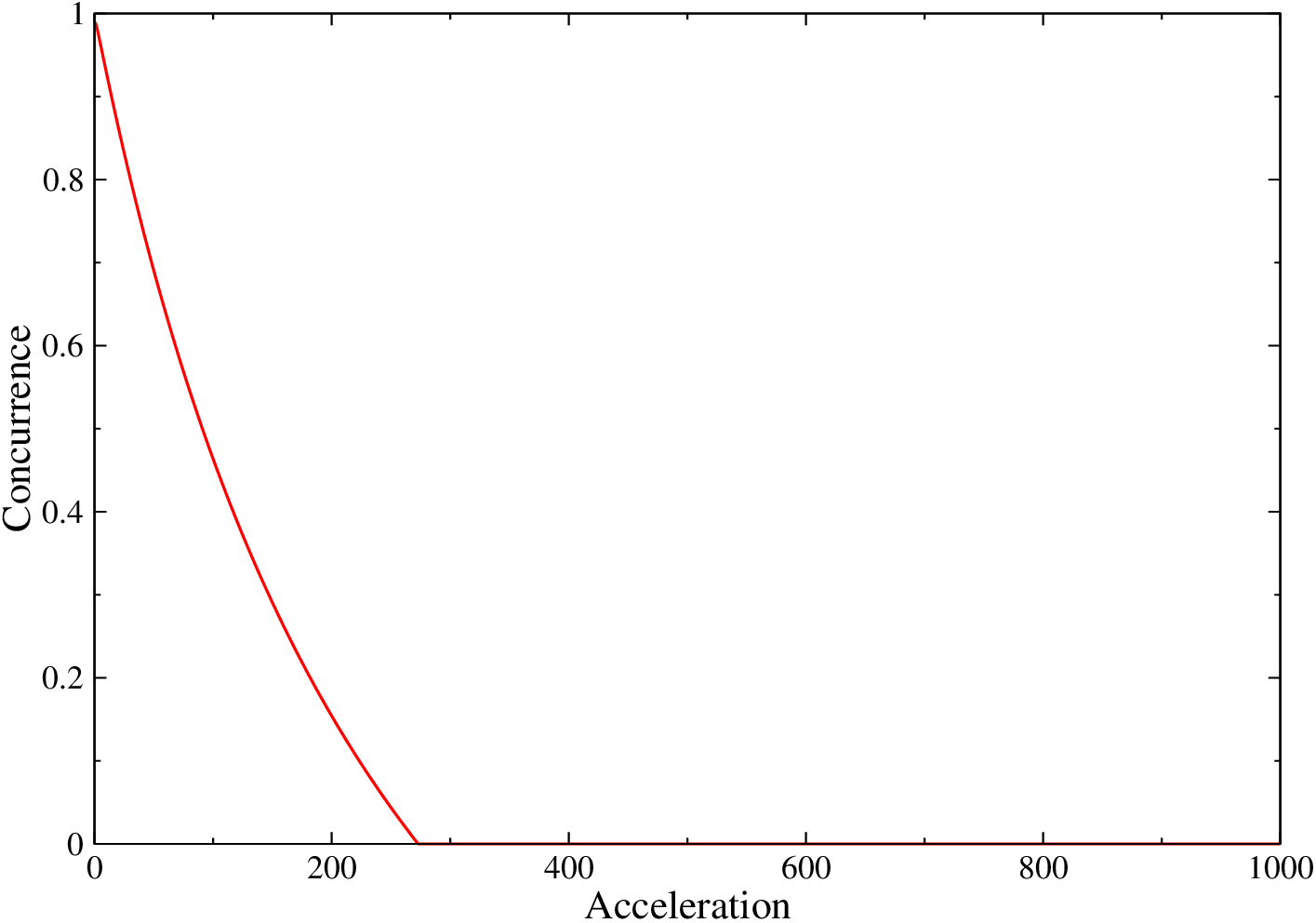}
\caption{The concurrence $C(\rho_{\infty}^{AR})$ is plotted as a function of the acceleration $a/\Omega$
assuming the same initial state and $\epsilon$, $\Omega$, $\Delta$, and $\kappa$ parameters
as in Fig.~\ref{Fig1_MI}. The sudden death of the entanglement between the two qubits is
observed at $a_{\rm sd}/\Omega \approx 273.00$.}
\label{Fig4_conc}
\end{center}
\end{figure}

Now, we show that the qubit-system entanglement experiences a sudden death
for accelerations smaller than the one necessary for the mutual information
to acquire its minimum. For this purpose, we calculate the concurrence~\cite{W98}
\begin{equation}
C(\rho_{\infty}^{AR})= \max \{0, \sqrt{\lambda_1} -\sqrt{\lambda_2} -\sqrt{\lambda_3} -\sqrt{\lambda_4}\},
\end{equation}
associated with our mixed state $\rho_\infty^{AR}$, where $\lambda_i $ $\;(i=1, \ldots, 4)$ are the
eigenvalues of
$\rho_{\infty}^{AR} (\sigma_y \otimes \sigma_y)  \overline{\rho}_{\infty}^{AR} (\sigma_y \otimes \sigma_y)$
with
$\lambda_1 \geq \lambda_2 \geq \lambda_3 \geq \lambda_4$
and
$\overline{\rho}_{\infty}^{AR}$ is obtained by taking the complex conjugate of every term in
Eq.~(\ref{estadofinal}).
In Fig.~\ref{Fig4_conc} we see that for arbitrarily small $a$, the qubit system has
$C(\rho_{\infty}^{AR}) \approx 1$ which is in agreement with $I(A:R) \approx 2$ found
in the low acceleration regime. Now, as the acceleration
increases the entanglement between the qubits decreases monotonically vanishing at a definite
value
$$
a/\Omega
=a_{\rm sd}/\Omega
= \pi/ \ln ( \nu^2/2 + \sqrt { 1+ \nu^4/4 \,} ).
$$
{\em Thus for a fixed acceleration time interval $\Delta$, the two qubits lose
their entanglement for every acceleration $a \geq a_{\rm sd}$.}


\section{Teleportation and the Unruh effect}
\label{sec:teleportation}

\begin{figure}[t]
\begin{center}
\includegraphics[height=0.22\textheight]{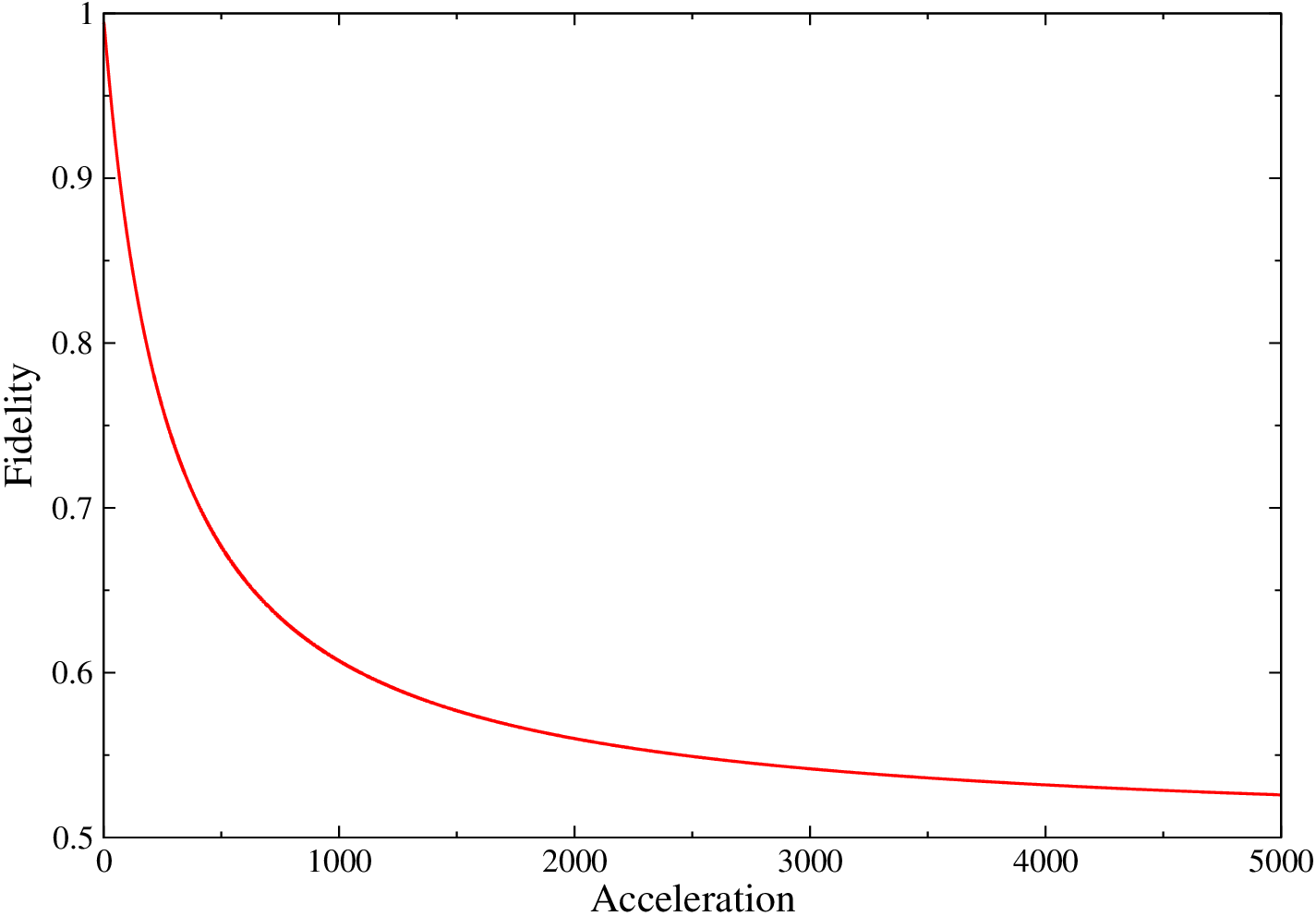}
\caption{The teleportation fidelity $F$ is plotted as a function of the acceleration
$a/\Omega$ with the values of $\epsilon$, $\Omega$, $\Delta$ and $\kappa$ being the
same as in Fig.~\ref{Fig1_MI}. }
\label{Fig5_fidelity}
\end{center}
\end{figure}

Now, let us use our previous results to revisit the teleportation protocol
when Alice and Rob initially share the entangled qubit system~(\ref{initialqubit})
in a singlet state, $\alpha=-\beta=1/\sqrt 2$, and calculate how the corresponding
fidelity is affected as a function of Rob's qubit acceleration. The state to be
teleported by Alice is given by
\begin{equation}
|\varphi_C \rangle= \gamma |0_C \rangle + \delta |1_C \rangle,
\label{phiC}
\end{equation}
which combined with $|\Psi_{AR} \rangle$ given in Eq.~(\ref{initialqubit})
and the Minkowski vacuum lead to the following total initial state
\begin{equation}
|\Psi^{CAR \phi}_{-\infty} \rangle
= |\varphi_C\rangle\otimes  |\Psi_{AR}\rangle\otimes |0_M\rangle.
\label{esttel}
\end{equation}
By using now that
\begin{eqnarray}
|0_C\rangle\otimes |0_A\rangle&=&\frac{1}{\sqrt{2}}(|\phi^+_{CA}\rangle + |\phi^-_{CA}\rangle),\nonumber \\
|0_C\rangle\otimes |1_A\rangle&=&\frac{1}{\sqrt{2}}(|\psi^+_{CA}\rangle + |\psi^-_{CA}\rangle),\nonumber \\
|1_C\rangle\otimes |0_A\rangle&=&\frac{1}{\sqrt{2}}(|\psi^+_{CA}\rangle - |\psi^-_{CA}\rangle), \nonumber \\
|1_C\rangle\otimes |1_A\rangle&=&\frac{1}{\sqrt{2}}(|\phi^+_{CA}\rangle - |\phi^-_{CA}\rangle),
\end{eqnarray}
where
$|\phi^+_{CA}\rangle, |\phi^-_{CA}\rangle, |\psi^+_{CA}\rangle , |\psi^-_{CA}\rangle$
are the Bell states~\cite{A07}, we cast Eq.~(\ref{esttel}) as
\begin{eqnarray}
|\Psi^{CAR \phi}_{- \infty} \rangle &=&
\frac{1}{2} [
     |\phi^+_{CA}\rangle\otimes (\gamma |1_R\rangle - \delta |0_R\rangle)\otimes|0_M\rangle \nonumber \\
      &&+ |\phi^-_{CA}\rangle\otimes (\gamma |1_R\rangle + \delta |0_R\rangle)\otimes|0_M\rangle \nonumber \\
      &&+ |\psi^+_{CA}\rangle\otimes (-\gamma |0_R\rangle + \delta |1_R\rangle)\otimes|0_M\rangle \nonumber \\
      &&- |\psi^-_{CA}\rangle\otimes (\gamma |0_R\rangle + \delta |1_R\rangle)\otimes|0_M\rangle
       ]. \nonumber
\end{eqnarray}
The asymptotic total final state after Rob has accelerated for proper time
$\Delta$ can be cast from Eq.~(\ref{evolution}) as
$$
|\Psi^{CAR \phi}_{\infty} \rangle
 =
(I + a_{R I}^{\dagger}(\lambda)R -  a_{R I}(\overline{\lambda}) R^{\dagger} )
|\Psi^{CAR \phi}_{-\infty} \rangle.
$$
For the sake of simplicity, let us assume that Alice makes a Bell measurement obtaining
$|\psi^- _{CA}\rangle$, which will be eventually informed to Rob by classical means.
Then, we have
\begin{eqnarray}
|\Psi^{CAR \phi}_{\infty} \rangle
& = &
-\frac{1}{2}|\psi^-_{CA}\rangle\otimes  (\gamma |0_R\rangle + \delta |1_R\rangle)\otimes |0_M\rangle
\nonumber \\
&+&
\frac{\gamma}{2}|\psi^-_{CA}\rangle\otimes |1_R\rangle \otimes a_{R I}(\overline{\lambda})|0_M\rangle
\nonumber \\
&-&
\frac{\delta}{2}|\psi^-_{CA}\rangle\otimes  |0_R\rangle \otimes a^{\dagger}_{RI}(\lambda)|0_M\rangle
\nonumber \\
&=&
-\frac{1}{2}|\psi^-_{CA}\rangle \otimes (\gamma |0_R\rangle + \delta |1_R\rangle)\otimes |0_M\rangle
\nonumber \\
&+&
\frac{\nu \gamma e^{-\pi \Omega/a}}{2 (1-e^{-2\pi \Omega/a})^{{1}/{2}}}|\psi^-_{CA}\rangle \otimes
|1_R \rangle \otimes|1_{\tilde{F}_{2\Omega}}\rangle
\nonumber \\
&-&
\frac{\nu \delta}{2 (1-e^{-2\pi \Omega/a})^{{1}/{2}}}|\psi^-_{CA}\rangle \otimes
|0_R \rangle \otimes |1_{\tilde{F}_{1\Omega}}\rangle.
\nonumber
\end{eqnarray}
The density matrix associated with Rob's qubit is
\begin{equation}
\rho_{\infty}^R =
||\Psi^{CAR \phi}_{\infty}||^{-2}\,
{\rm tr}_{\phi CA} |\Psi^{CAR \phi}_{\infty}\rangle \langle \Psi^{CAR \phi}_{\infty}|,
\label{rho_R1}
\end{equation}
where
\begin{equation}
||\Psi^{CAR \phi}_{\infty}||^2 =
\frac{1}{4}\left(
1+ \frac{|\gamma|^2 \nu^2 e^{-2\pi \Omega/a}}{1-e^{-2\pi \Omega/a}}
+ \frac{|\delta|^2 \nu^2}{1-e^{-2\pi \Omega/a}}
           \right).
\end{equation}
Eq.~(\ref{rho_R1}) can be recast as
\begin{eqnarray}
\rho_{\infty}^R
& = &
(|\gamma|^2\, S_0^{\gamma \delta} +
 |\delta|^2\, S_2^{\gamma \delta}) |0_R\rangle\langle0_R|
+ \gamma\, \overline{\delta}\, S_0^{\gamma \delta} |0_R\rangle\langle1_R| \nonumber \\
& + & \overline{\gamma}\, \delta\,  S_0^{\gamma \delta} |1_R\rangle\langle0_R|
+(|\delta|^2\, S_0^{\gamma \delta}
+ |\gamma|^2\, S_1^{\gamma \delta}) |1_R\rangle\langle1_R|,\nonumber \\
\end{eqnarray}
where
\begin{eqnarray}
 S^{\gamma \delta}_0 = \frac{1-e^{-2\pi\Omega/a}}{1-e^{-2\pi\Omega/a}  + \nu^2|\delta|^2 + \nu^2|\gamma|^2
 e^{-2\pi\Omega/a}}, \\
 S^{\gamma \delta}_1 = \frac{\nu^2e^{-2\pi\Omega/a}}{1-e^{-2\pi\Omega/a}  + \nu^2|\delta|^2 +
 \nu^2|\gamma|^2e^{-2\pi\Omega/a}}, \\
 S^{\gamma \delta}_2 =\frac{\nu^2}{1-e^{-2\pi\Omega/a} + \nu^2|\delta|^2 + \nu^2|\gamma|^2 e^{-2\pi\Omega/a}}.
\end{eqnarray}
Let us choose $\gamma= \delta = 1/\sqrt{2}$ in Eq.~(\ref{phiC}).
In this case, using the basis $ \{|0_R\rangle, |1_R\rangle \}$
we have
\begin{equation}
\rho_{\infty}^R=\left( \begin{array}{c c}
S_0+ S_2 & S_0 \\
S_0 & S_0 +S_1
\end{array}
\right),
\label{estadofinaltel}
\end{equation}
with
$$
S_0 \equiv \frac{S_0^{1/ \sqrt{2}\; 1/\sqrt{2}}}{2},\,
S_1 \equiv \frac{S_1^{1/ \sqrt{2}\; 1/\sqrt{2}}}{2},\,
S_2 \equiv \frac{S_2^{1/ \sqrt{2}\; 1/\sqrt{2}}}{2}.
$$
Finally, the teleportation fidelity
$F \equiv \langle\varphi_C |\rho_{\infty}^R| \varphi_C \rangle$
turns out to be
\begin{equation}
F = S_0 + {1}/{2},
\end{equation}
which is plotted in Fig.~\ref{Fig5_fidelity} as a function of Rob's qubit proper
acceleration.
We see from Fig~\ref{Fig5_fidelity} that for low enough accelerations $F\approx 1$ and
for arbitrarily large accelerations $F\approx 0.5$. This is so because of the
entanglement loss between Alice and Rob's qubits as discussed in the previous
section. In contrast to Figs.~\ref{Fig1_MI} and~\ref{Fig2_ent} we see that $F$ has
a monotonous decrease as a function of $a/\Omega$.

\section{Final remarks}
\label{sec:conclusions}

Technological developments have recently provided means to new and
exquisite tests of quantum mechanics. This is not only interesting in
connection with information theory but also with a number of conceptual
issues. In particular the interplay between quantum mechanics and relativity
has been a permanent source of preoccupation~\cite{PT04} which culminates
with the long standing quest for quantum gravity. However, very
interesting physics involving quantum mechanics and relativity can be
already witnessed in Minkowski spacetime as, e.g., the fact that
spin entanglement and entropy are not invariant by Lorentz
transformations when the associated particles are described by wave
packets~\cite{PST02,GA02}. A consequence coming out from these facts
is that Bell inequalities  can be {\em satisfied} rather than violated if
the spin detectors move fast enough~\cite{LM09}. In the present paper,
we have analyzed how the teleportation fidelity
is affected when one of the entangled qubits is uniformly accelerated
for a finite time interval under the influence of some external agent.
We model our qubit to interact with a massless scalar field
as it accelerates. An hypothetical laboratory realization of our model
can be envisaged by using as qubit a charged fermion
accelerated by an electric field pointing in the same direction of
some background magnetic field along which the fermion spin is
prepared~\cite{DH09}. The coupling between the spin and magnetic field
gives rise to the qubit internal energy gap. Then, the unexcited and excited
qubit states correspond to the cases where the spin points in the same
and opposite directions with respect to the magnetic field, respectively.
We have shown that the teleportation fidelity steadily decays as the
acceleration increases for a fixed interaction proper time
(see Fig.~\ref{Fig5_fidelity}). From the point of view of inertial observers this is
due to the fact that part of the entanglement between the qubits is carried
away by the scalar radiation which is emitted when the accelerated
qubit suffers a transition. This is confirmed by the fact that the
{\em qubit-system mutual information} and the
{\em qubit system-field entanglement} have a complementary behavior
as a function of the acceleration magnitude, i.e. one decreases (increases)
as the other one increases (decreases) (see Figs.~\ref{Fig1_MI}
and~\ref{Fig2_ent}). The non-triviality of these graphs, codified by the
fact that the lines do not have a monotonous behavior can be
understood from Fig.~\ref{Fig3_time}, which shows that after
some long enough time $\tau_e$ the entanglement between
the qubit system and field begins to decrease back.
This is obvious for the case $a =2000$. For $a \leq 545.75$ this behavior
would also be seen if $\Delta$ were large enough. Remarkably, the
concurrence which measures the entanglement of the qubit system
experiences a sudden death for some acceleration
$a_{\rm sd}$ as shown in Fig.~\ref{Fig4_conc}. Finally, this is in order to
call attention that from the point of view of uniformly accelerated
observers the interpretation for the above results is quite different
from the one due to inertial observers, since from their point of view
the uniformly accelerated qubit interacts with the very Unruh thermal
bath of real (Rindler) particles in which it is immersed in its proper
frame. This is another example of how inertial and accelerated
observers can give quite different physical interpretations concerning
the same physical phenomenon although they must of course agree on the
output measured by a given experimental set up (see, e.g.,
Ref.~\cite{VM01}).

\acknowledgments

A.L. and G.M. acknowledge full and partial financial support from
Funda\c c\~ao de Amparo \`a Pesquisa do Estado de S\~ao Paulo,
respectively. G.M. Also acknowledges partial support from Conselho
Nacional de Desenvolvimento Cient\'\i fico e Tecnol\'ogico.

\appendix
\section{Derivation of Eq.~(\ref{nu})}
\label{appendix}

Here we calculate the $\nu$ coefficient introduced in Eq.~(\ref{nu}).
For this purpose, let us consider a general smooth compact support
function $f \in C_0^{\infty}(M)$ defined in a globally hyperbolic
time-orientable spacetime $(M, g_{a b})$ and choose a Cauchy surface
$\Sigma \subset  M- J^{+}({\rm supp}\; f)$ outside the causal future
of its support~\cite{wald90}.
Now, let us define
\begin{equation}
\lambda (x) \equiv \int_M  d^4x'\sqrt{-g(x') } G^{\rm adv}(x, x') \, f(x').
\end{equation}
Then, $(\nabla^a\nabla_a- m^2)\lambda= f$ and we note that
${\rm supp}\; \lambda \subset J^-({\rm supp}\; f)$. Hence,
assuming $\phi \in C^{\infty}(M)$ to be any solution of
Eq.~(\ref{KG}), we have
\begin{eqnarray}
\int_{M} d^4x \sqrt{-g} \, \phi  \, f
&=&\int_{J^+(\Sigma)}d^4x \sqrt{-g} \,  \phi  \, f
\nonumber \\
&=&\int_{J^+(\Sigma)} d^4x \sqrt{-g} \,  \phi  \, (\nabla^a\nabla_a- m^2)\lambda
\nonumber \\
&=&\int_{J^+(\Sigma)} d^4x \sqrt{-g}  \, \nabla^a(\phi\nabla_a\lambda - \lambda \nabla_a\phi)
\nonumber\\
&+&\int_{J^+(\Sigma)} d^4x \sqrt{-g} \,  \lambda \, (\nabla^a\nabla_a- m^2)\phi
\nonumber \\
&=&\int_{\Sigma} d^3x \sqrt{h}  \, (\phi\nabla_a\lambda - \lambda\nabla_a\phi)n^a,
\nonumber
\end{eqnarray}
where
$n^a$ is a unit normal vector orthogonal to $\Sigma$. Now, by using
Eq.~(\ref{Ef}), we see that $Ef|_{\Sigma}=\lambda|_{\Sigma}$ and thus
\begin{equation}
\int_{M} d^4x  \sqrt{-g} \, \phi \,  f  = \int_{\Sigma} d^3x \sqrt{h} \,  (\phi\nabla_a (Ef)
                           - (Ef) \nabla_a\phi) n^a.
\label{lema}
\end{equation}
Next, let us decompose $Ef$ in terms of positive- and negative-frequency
Rindler modes
$u_{\omega \mathbf{k}_{\bot}}$
and
$\overline{u}_{\omega \mathbf{k}_{\bot}}$,
respectively, as
\begin{eqnarray}
Ef
& = &
\int_{0}^{\infty} d\omega\int d\mathbf{k}_{\bot}
[(u_{\omega \mathbf{k}_{\bot}}, Ef)_{KG} u_{\omega \mathbf{k}_{\bot}}
\nonumber \\
& - &
(\overline{u}_{\omega \mathbf{k}_{\bot}}, Ef)_{KG}
\overline{u}_{\omega \mathbf{k}_{\bot}}],
\label{Ef2}
\end{eqnarray}
where
$u_{\omega \mathbf{k}_{\bot}}$
satisfies
$\nabla_a \nabla^a u_{\omega \mathbf{k}_{\bot}}=0 $
with
$\mathbf{k}_\bot \equiv (k_y, k_z)$
and
is eigenstate of $i\partial_\tau$,
$-i\partial_y$ and $-i\partial_z$ with eigenvalues $\omega$,
$k_y$ and $k_z$, respectively.
Then, from Eq.~(\ref{lema}) we have
\begin{eqnarray}
(u_{\omega \mathbf{k}_{\bot}}, Ef)_{KG}
&=&
i \int_M  d^4x \sqrt{-g}  \, f  \, \overline{u}_{\omega \mathbf{k}_{\bot}} ,
\label{pos}\\
(\overline{u}_{\omega \mathbf{k}_{\bot}}, Ef)_{KG}
&=&
i\int_M d^4x \sqrt{-g}  \, f  \, u_{\omega \mathbf{k}_{\bot}}.
\label{neg}
\end{eqnarray}
Let us now show that Eq.~(\ref{neg}) vanishes. For this purpose, we write
$u_{\omega \mathbf{k}_{\bot}}=e^{-i\omega\tau}\varphi_{\omega \mathbf{k}_{\bot}}(\xi, {\bf x}_\bot)$,
where
$$
\varphi_{\omega \mathbf{k_{\bot}}}(\xi, {\bf x}_\bot)
=
\left[\frac{\sinh{({\pi \omega}/{a})}}{4\pi^4 a}
\right]^{{1}/{2}}K_{i{\omega}/{a}}({k_{\bot}}e^{a\xi}/a)
e^{i \mathbf{k}_{\bot}\cdot \mathbf{x}_\bot}
$$
with $K_\mu (z)$ being the modified Bessel function,
$\mathbf{x}_\bot \equiv (y,z)$
and we are covering the right Rindler wedge with coordinates
$(\tau, \xi, {\bf x}_\bot)$ in which case the corresponding
line element becomes
$$
ds^2 = e^{2a\xi} (- d\tau^2 + d\xi^2) + d{\bf x}^2_\bot.
$$
Then, we integrate Eq.~(\ref{neg}) in the $\tau$
variable by using $\epsilon(\tau)\approx \epsilon = {\rm const}$
when the detector is switched on (and $\epsilon(\tau) = 0$ when the
detector is switched off), obtaining
\begin{equation}
(\overline{u}_{\omega \mathbf{k}_{\bot}},Ef)_{KG}
=
2i\epsilon\gamma_{\omega \mathbf{k}_{\bot}}
\frac{\sin{[(\omega+\Omega){\Delta}/{2}]}}{(\omega+\Omega)},
\label{aux2}
\end{equation}
where
$
\gamma_{\omega \mathbf{k}_{\bot}}
\equiv
\int_{\Sigma} d^3x \sqrt{-g} \psi({\bf x})\varphi_{\omega \mathbf{k}_{\bot}}
$.
Then, by using the fact that
$$
\frac{\sin{[(\omega+\Omega){\Delta}/{2}]}}{(\omega+\Omega)} \approx \pi \delta (\omega+\Omega)
$$
when $\Delta  \gg \Omega^{-1}$, we have
$(\overline{u}_{\omega \mathbf{k}_{\bot}},Ef)_{KG} \approx 0$.
Thus
$Ef$ is approximately a positive-frequency solution, i.e., $KEf\approx Ef$.
An analogous reasoning can be used to show that $E\overline{f}$ is a
negative-frequency solution, i.e.,  $KE\overline{f}\approx 0$.
Analogously to Eq.~(\ref{aux2}), we have
\begin{equation}
({u}_{\omega \mathbf{k}_{\bot}}, Ef)_{KG}
=
2 i\epsilon \gamma_{\omega \mathbf{k}_{\bot}}
\frac{\sin{[(\omega-\Omega){\Delta}/{2}]}}{(\omega-\Omega)}.
\end{equation}
Now, by using Eqs.~(\ref{nulambda}) and~(\ref{Ef2}), we write
\begin{eqnarray}
\nu^2
&\equiv& ||\lambda||^2  = ||KEf||^2
\nonumber \\
& = & \int_{0}^{\infty} d\omega\int d\mathbf{k}_{\bot}
|({u}_{\omega \mathbf{k}_{\bot}}, Ef)_{KG}|^2
\nonumber \\
&\approx &
2\pi \epsilon^2 \Delta \int d\mathbf{k}_{\bot} |\gamma_{\Omega \mathbf{k}_{\bot}}|^2.
\label{acelerado2}
\end{eqnarray}
In the particular case where we have a point detector,
$\psi({\bf x}) \to \delta({\bf x})$, we end up with
\begin{equation}
\nu^2 =\frac{\epsilon^2 \Omega\Delta}{2\pi}.
\label{acelerado}
\end{equation}
For small but not point detectors, let us calculate $\nu$  assuming
$$
\psi(\mathbf{x})=\frac{e^{-\mathbf{x}^2/2\kappa^2}}{(\kappa\sqrt{2\pi})^3}
$$
in the inertial case, where $\kappa = {\rm const}$ is the Gaussian variance. Then,
\begin{eqnarray}
\nu^2_{\rm in}
&=&
\int d\mathbf{k}|(v_{\mathbf{k}}, Ef)_{KG}|^2
\nonumber \\
&\approx&
\frac{\epsilon^2}{4\pi}
\int d\mathbf{k}
\frac{ \delta(\omega_{\mathbf{k}}-\Omega)}{\omega_{\mathbf{k}}}
\frac{\sin{[(\omega_{\mathbf{k}}}-\Omega){\Delta}/{2}]}{(\omega_{\mathbf{k}} -\Omega)}
|\hat{\psi}(-\mathbf{k})|^2
\nonumber
\end{eqnarray}
with
$
v_{\mathbf{k}} = {e^{i({\bf k}{\bf x} - \omega t})}/{\sqrt{16 \pi^3 \omega_{\bf k}}}
$
being positive-frequency Minkowski modes and
$\hat{\psi}(\mathbf{k})$ the Fourier transform of
$\psi(\mathbf{x})$.
Finally, by using $\omega_{\mathbf{k}}=|\mathbf{k}|$ and integrating in spherical coordinates
we find
\begin{equation}
\nu^2_{\rm in}=\frac{\epsilon^2 \Omega\Delta}{2\pi}e^{-\Omega^2 \kappa^2}.
\label{inercial}
\end{equation}
Because in the point detector case, $\kappa=0$, Eqs.~(\ref{acelerado})
and~(\ref{inercial}) are identical, we shall use Eq.~(\ref{inercial})
as an approximation for Eq.~(\ref{acelerado2}) associated with the accelerated
case provided that $\kappa \ll 1$. This drives us to Eq.~(\ref{nu}).

\end{document}